\documentclass[12pt,preprint]{aastex}

\usepackage{color,txfonts}

\begin{document}

\title{{\it Fermi} Large Area Telescope detection of gamma-ray emission 
from the direction of supernova iPTF14hls}

\author{Qiang Yuan$^{1,2}$, Neng-Hui Liao$^1$, Yu-Liang Xin$^{1,3}$,
Ye Li$^4$, Yi-Zhong Fan$^{1,2}$, Bing Zhang$^5$, Hong-Bo Hu$^{6,3}$,
Xiao-Jun Bi$^{6,3}$}

\affil{$^1$Key laboratory of Dark Matter and Space Astronomy, Purple
Mountain Observatory, Chinese Academy of Sciences, Nanjing 210008, China;
yuanq@pmo.ac.cn, yzfan@pmo.ac.cn\\
$^2$School of Astronomy and Space Science, University of Science and
Technology of China, Hefei 230026, Anhui, China\\
$^3$University of Chinese Academy of Sciences, 19A Yuquan Road, Beijing
100049, China\\
$^4$Kavli Institute of Astronomy and Astrophysics, Peking University,
Beijing 100871, China\\
$^5$Department of Physics and Astronomy, University of Nevada Las Vegas,
Las Vegas, NV 89154, USA; zhang@physics.unlv.edu\\
$^6$Key Laboratory of Particle Astrophysics, Institute of High Energy
Physics, Chinese Academy of Sciences, Beijing 100049, China;
huhb@ihep.ac.cn
}

\begin{abstract}

The remnant of supernova explosion is widely believed to be the acceleration
site of high-energy cosmic ray particles. The acceleration timescale is,
however, typically very long. Here we report the detection of a variable
$\gamma$-ray source with the {\it Fermi} Large Area Telescope, which is
positionally and temporally consistent with a peculiar supernova,
iPTF14hls. A quasi-stellar object SDSS J092054.04+504251.5, which is
probably a blazar candidate according to the infrared data, is found
in the error circle of the $\gamma$-ray source. More data about the
$\gamma$-ray source and SDSS J092054.04+504251.5 are needed to confirm
their association. On the other hand, if the association between the
$\gamma$-ray source and the supernova is confirmed, this would be the 
first time to detect high-energy $\gamma$-ray emission from a supernova, 
suggesting very fast particle acceleration by supernova explosions. 

\end{abstract}


\section{Introduction}

Observations of $\gamma$-ray emission from supernova remnants (SNRs) 
prove that they are high-energy cosmic ray accelerators
\citep{2004Natur.432...75A,2013Sci...339..807A}. Particle
acceleration typically occurs in a very long timescale, e.g., hundreds
to thousands of years, after the supernova explosion. This picture was
consistent with the non-detection of $\gamma$-ray emission from the most
nearby supernova, 1987A~\citep{1988MNRAS.234P..73S,1996ApJ...472..800Y,
2003ApJ...591L..25E}. It has been expected that fast particle
acceleration might occur soon after supernova explosion if there
were interactions between the ejecta and pre-existing dense material
\citep{2011PhRvD..84d3003M} or there was a young, powerful pulsar wind
nebula \citep{2015ApJ...805...82M}. However, previous searches for 
$\gamma$-ray emission from supernovae located in dense circumstellar 
medium \citep{2015ApJ...807..169A} or from super-luminous supernovae 
\citep{2017arXiv170808971R} with the {\it Fermi} Large Area Telescope 
\citep[{\it Fermi}-LAT;][]{2009ApJ...697.1071A} have not led to a positive
detection.

A very peculiar supernova, iPTF14hls, was discovered by the
Intermediate Palomar Transient Factory on 2014 September 22.53 UT
\citep{2017Natur.551..210A}. This event has the identical spectra 
compared with a typical hydrogen-rich core-collapse supernova, but shows 
a very different light curve. The optical emission from iPTF14hls 
remains bright up to 600 days after the first detection, and experiences 
at least five rebrightenings in two years. What is even more surprising 
is that there was possibly an eruption $\sim 60$ years ago in 1954
at the same position of iPTF14hls. This event challenges the
traditional understanding of the explosions of massive stars at the 
end of their lives \citep{2017Natur.551..173W}.

Multi-wavelength observations are particularly important for
understanding the nature of such a peculiar event. The observation
in the X-ray band by Swift/XRT on 2015 May 23.05, which was about 244
days after the discovery, showed no detection of the source and gave
an upper limit of luminosity of $L_X<2.5\times10^{41}$ erg~s$^{-1}$
in the $0.3-10.0$ keV band assuming a photon index of $\Gamma=2$
and a neutral hydrogen column density of $1.4\times10^{20}$ cm$^{-2}$
\citep{2017Natur.551..210A}. In 2015 May and 2016 June, the source
was observed by the Arcminute Microkelvin Imager Large Array and the
Very Large Array in the radio bands, resulting in no detection in all 
those observations \citep{2017Natur.551..210A}. These observations 
show evidence against interaction between the supernova ejecta and
pre-existing material. However, a late time observation after $\sim3$ 
years of the explosion revealed strong interaction between the 
shock and the circumstellar material \citep{2017arXiv171200514A}.

Here we report the search for potential emission from iPTF14hls in the
$\gamma$-ray band, using the data of the {\it Fermi}-LAT. We find a variable
$\gamma$-ray source which is potentially coincident with iPTF14hls.
This could be the first time to detect high-energy $\gamma$-ray
emission from a supernova, if the association is true.
However, the interpretation of the $\gamma$-ray emission seems to
be difficult under the traditional scenario of supernova explosion,
given the constraints from the X-ray and radio observations.
On the other hand, we find a quasar within the error circle of the
{\it Fermi}-LAT source, which is potentially a blazar candidate according 
to the infrared color diagram. No radio counterpart is found for this
quasar, making its association with the $\gamma$-ray source still
uncertain. More observations are necessary to firmly address the
association of the $\gamma$-ray source with either the supernova
or the blazar candidate (e.g., follow-up multi-wavelength 
monitorings of the candidate blazar and continuous {\it Fermi}-LAT 
observations of the $\gamma$-ray source).

\section{{\it Fermi}-LAT observations}

We use {\it Fermi}-LAT data recorded from 2008 August 4 to 2017 November 
16, restricted to the Pass 8 Source class (evclass $=128$ \& evtype $=3$).
We select the data in a $14^\circ \times 14^\circ$ box
region centered at the target source iPTF14hls with energies between
200 MeV\footnote{The 200 MeV lower threshold enables us to have a 
better angular resolution of selected events. Furthermore, as can be seen
below, the $\gamma$-ray spectrum of this source is relatively hard, and
there is not much emission at low energies. We have tested that choosing
a 100 MeV threshold gives similar results.} and 500 GeV. The events with 
zenith angles $>90^\circ$ are excluded to reduce the contamination from 
the Earth Limb. We use the {\it Fermi}-LAT Science Tools 
v10r0p5\footnote{http://fermi.gsfc.nasa.gov/ssc/data/analysis/software/} 
and the standard binned likelihood analysis method {\tt gtlike} to analyze 
the data. The sources in the third {\it Fermi}-LAT catalog \citep[3FGL;][]
{2015ApJS..218...23A}, together with the diffuse Galactic and isotropic 
backgrounds\footnote{http://fermi.gsfc.nasa.gov/ssc/data/access/lat/BackgroundModels.html} 
{\tt gll\_iem\_v06.fits} and {\tt iso\_P8R2\_SOURCE\_V6\_v06.txt}, are 
included in the model. An additional field source at position 
$({\rm R.A.,Dec.})=(143.12^{\circ},53.08^{\circ})$ has also been added, 
according to the residual map of the fitting.

\begin{figure*}[!htb]
\centering
\includegraphics[width=0.32\textwidth]{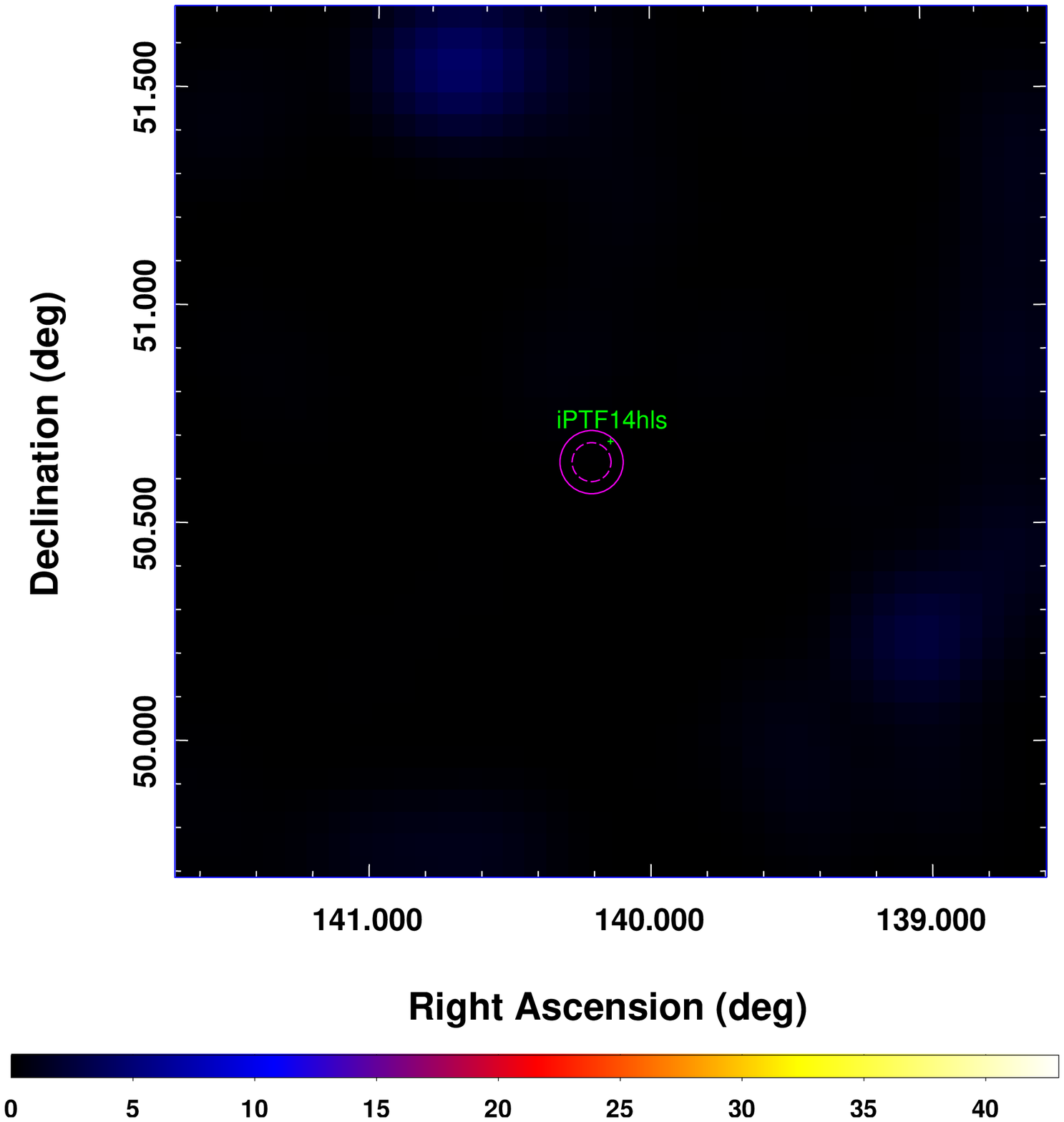}
\includegraphics[width=0.32\textwidth]{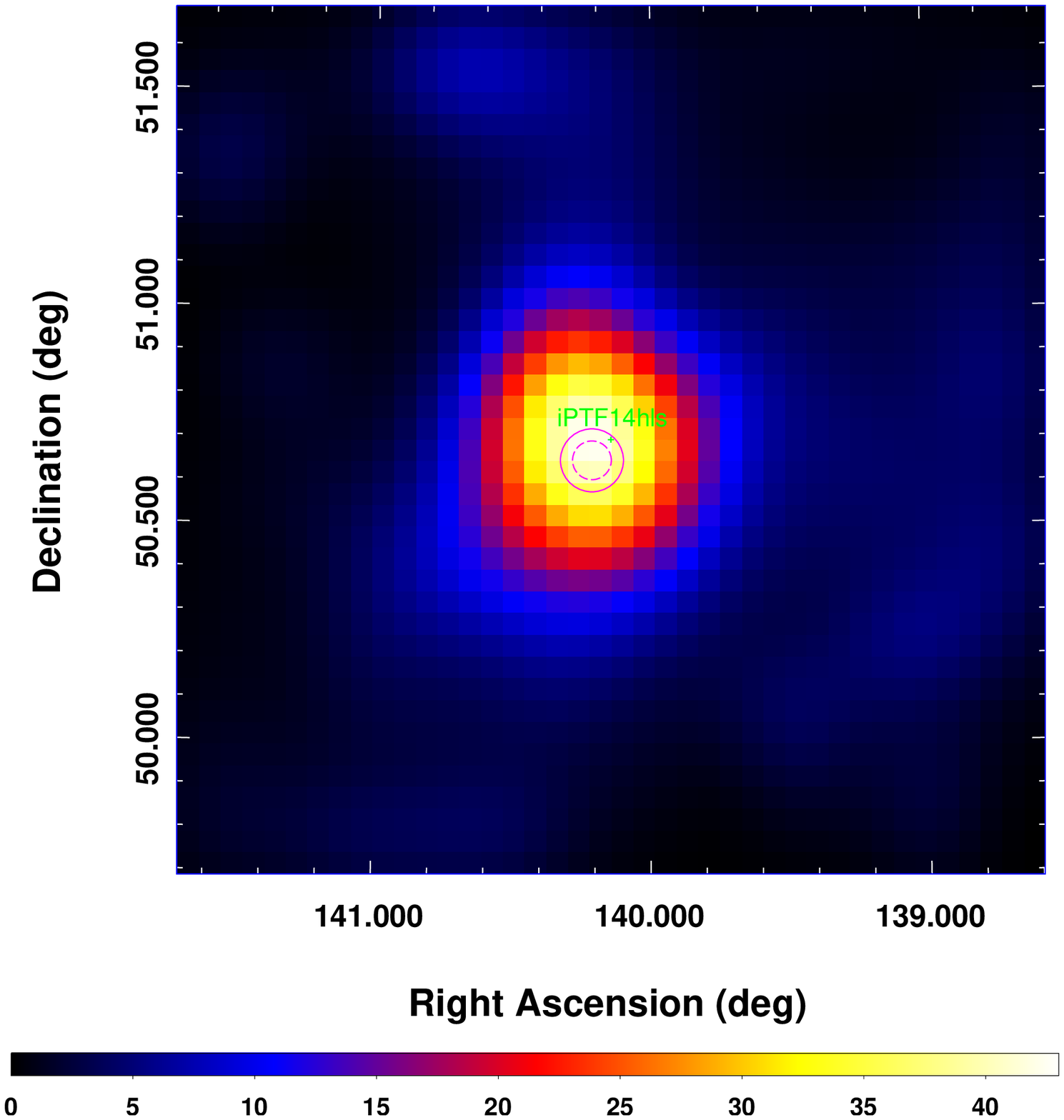}
\includegraphics[width=0.32\textwidth]{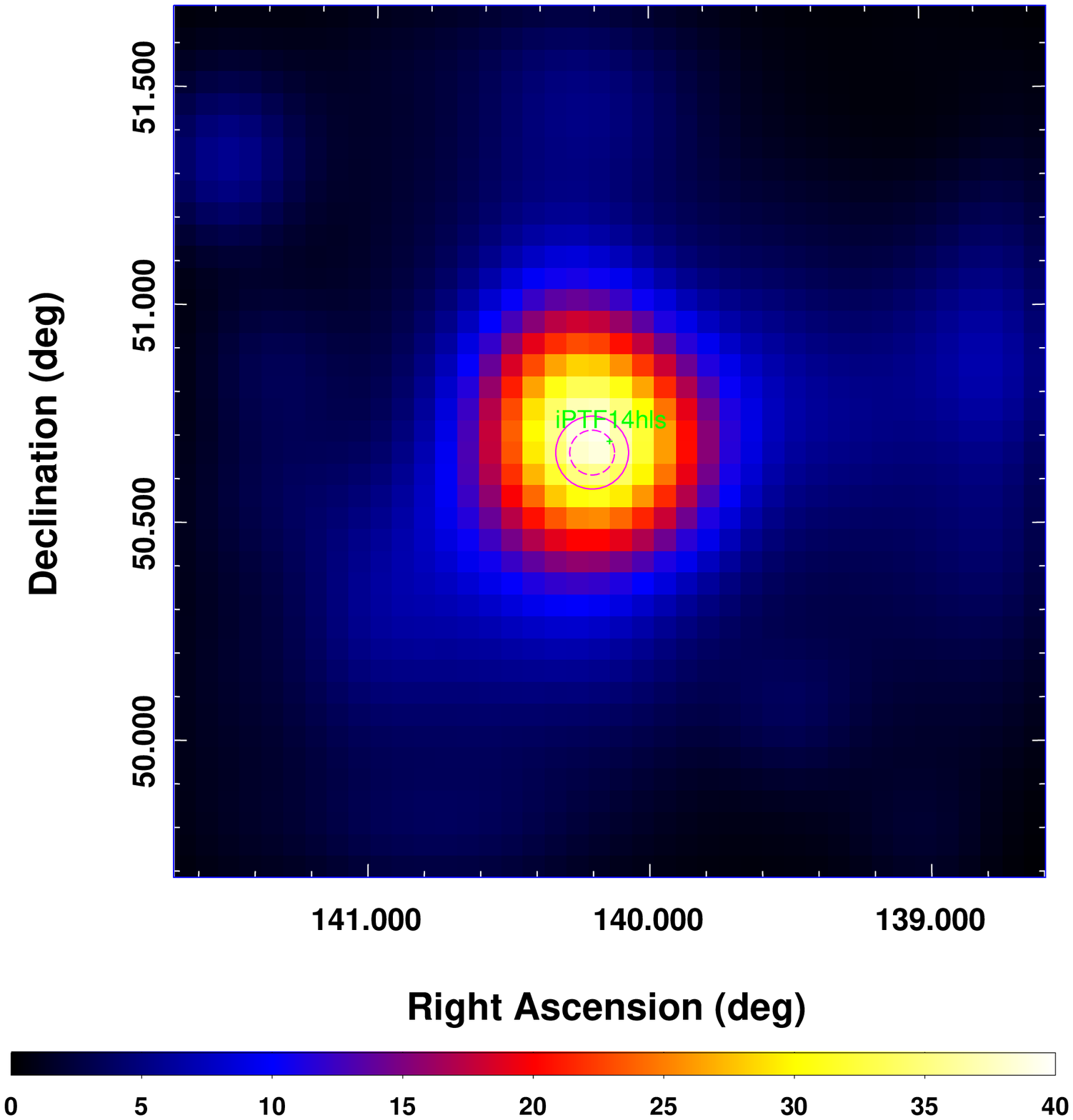}
\caption{TS maps for a $2^{\circ}\times 2^{\circ}$ region centered
at iPTF14hls, for the {\it Fermi}-LAT data before (left) and after (middle)
2014 September 22. The right panel shows the TS map for the past two
years of data. All the maps are smoothed with a Gaussian kernel with a
width of $0.2^{\circ}$. Magenta circles in these plots show the 68\% 
(inner) and 95\% (outer) error regions of the $\gamma$-ray localization.
}
\label{fig:tsmap}
\end{figure*}

We divide the data into two parts, before and after the explosion
date of iPTF14hls, and perform the likelihood analysis, respectively.
For the $\sim6$ years of data before 2014 September 22, we find no
significant emission at the position of iPTF14hls. The Test Statistic
(TS) map for a $2^{\circ}\times 2^{\circ}$ region centered at iPTF14hls
is shown in the left panel of Fig.~\ref{fig:tsmap}. For the data after
the supernova explosion, a clear $\gamma$-ray source appears in the data, 
as can be seen in the middle panel of
Fig.~\ref{fig:tsmap} for the TS map assuming the same model as that
used in the left one. Assuming a power-law point source right at the
location of iPTF14hls, we find a TS value of $\sim53$, and a spectral
index of $2.03\pm0.16$. The best-fit position of the source using
{\tt gtfindsrc} turns out to be $({\rm R.A.,Dec.})=(140.21^{\circ},
50.65^{\circ})$, with a 68\% (95\%) error circle of $0.045^{\circ}$ 
($0.073^{\circ}$). The distance between iPTF14hls and the best-fit 
position of the $\gamma$-ray source is about $0.065^{\circ}$. Therefore 
iPTF14hls is positionally consistent with the {\it Fermi}-LAT variable 
source at the 95\% confidence level. We also perform an analysis of the 
data from 2015 August 4 to 2017 November 16, i.e., about one year after
the explosion date (see below for the light curve analysis),
which results in a best-fit position of $({\rm R.A.,Dec.})=
(140.21^{\circ}, 50.67^{\circ})$, and a 68\% error circle radius of
$0.051^{\circ}$. The separation between iPTF14hls and the best-fit 
position is $0.048^{\circ}$, within the $68\%$ error circle (see the 
right panel of Fig.~\ref{fig:tsmap}). In the following analysis, we will 
present the results based on the past three years of data (from 2014 
September 22 to 2017 November 16), unless stated explicitly.

The $\gamma$-ray flux of the source is about $1.5\times10^{-9}$
cm$^{-2}$~s$^{-1}$ between 0.2 and 500 GeV. For a distance of 156 Mpc
\citep{2017Natur.551..210A}, it corresponds to a $\gamma$-ray luminosity 
of $1.0\times10^{43}$ erg~s$^{-1}$. This value is comparable with the 
peak bolometric luminosity of iPTF14hls \citep{2017Natur.551..210A}. 
Assuming a 3 year emission time, the total energy released in 
$\gamma$-rays is estimated to be about $10^{51}$ erg.

\begin{figure*}[!htb]
\centering
\includegraphics[width=0.48\textwidth]{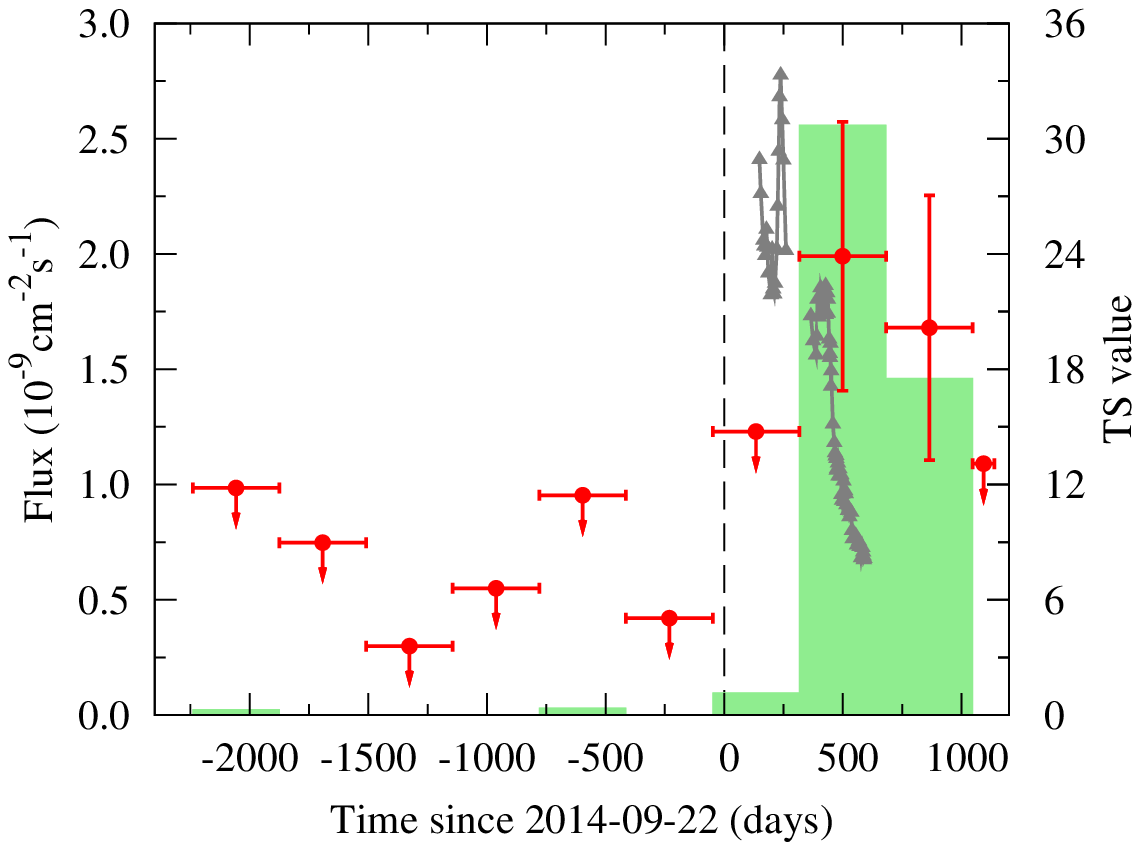}
\includegraphics[width=0.48\textwidth]{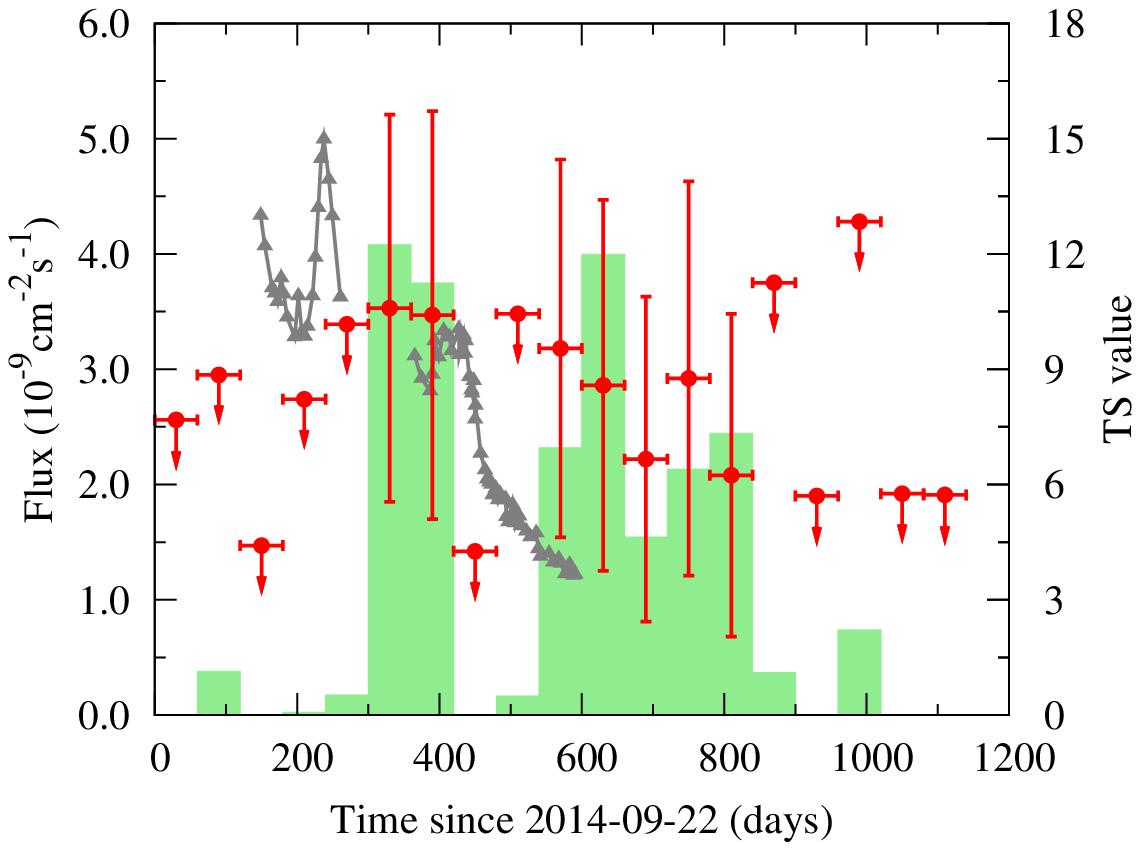}
\caption{Light curves of $\gamma$-ray emission from the direction of
iPTF14hls between 0.2 and 500 GeV in 1-year (left) and 2-month (right)
bins. The zero point is adopted to be September 22, 2014. Shaded regions
show the TS values (right axis). For TS values less than 4, the $95\%$
flux upper limits are presented. The bolometric luminosity with an
arbitrary normalization of iPTF14hls deduced from the blackbody fits 
\citep{2017Natur.551..210A} is shown by gray triangles for comparison.
}
\label{fig:lc}
\end{figure*}

\begin{figure*}[!htb]
\centering
\includegraphics[width=0.7\textwidth]{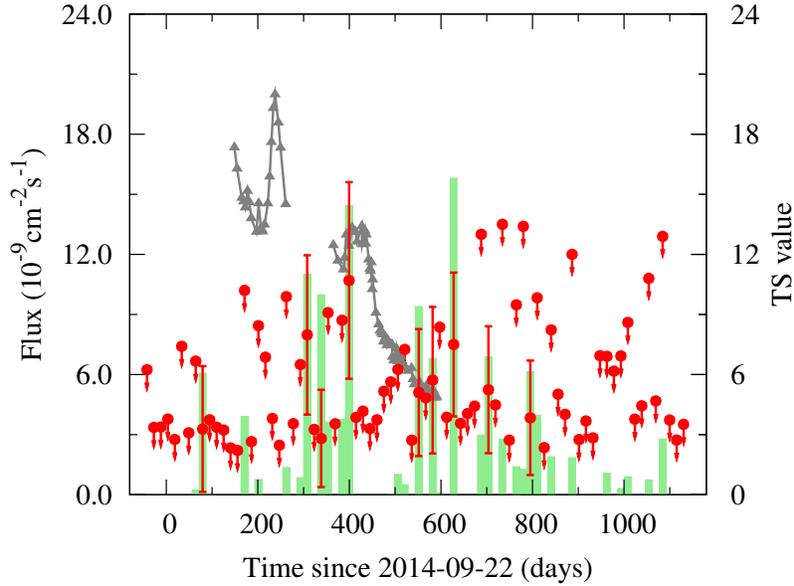}
\caption{Same as Fig.~\ref{fig:lc} but for 2-week bins of the light curve.
}
\label{fig:lc2w}
\end{figure*}

To further address the variabilities of the $\gamma$-ray emission
of the source, we derive the light curves of the $\gamma$-ray fluxes
between 0.2 and 500 GeV, for 1-year and 2-month bins, which are shown
in Fig.~\ref{fig:lc}. For the time bins in which the TS values are
smaller than 4, the 95\% flux upper limits are given. We find that
the source starts to emit $\gamma$-rays about 300 days after the
explosion time of iPTF14hls, and the emission lasts for about 850
days. Weak emission may last for even longer time, but the significance
becomes too low. We estimate the TS value of variabilities 
via ${\rm TS}_{\rm var}=-2\sum_i\ln\left[{\mathcal L}_i(F_{\rm const})/
{\mathcal L}_i(F_i)\right]$, where ${\mathcal L}_i(F_{\rm const})$ and
${\mathcal L}_i(F_i)$ are the likelihoods in the $i$th time bin for
constant and variable flux assumptions~\citep{2012ApJS..199...31N}. 
The TS value of the variability during the period of 300 and 850 days 
after the supernova explosion is about 11.4 for the 2-month binning
light curve, which roughly corresponds to a $1.3\sigma$ significance
for 8 more degrees of freedom of the variable hypothesis. 
The emission is probably variable at even shorter timescales 
(see Fig.~\ref{fig:lc2w} for the light curve for 2-week bins). However, 
for most bins the TS values are too small to draw a clear conclusion.

\begin{figure*}[!htb]
\centering
\includegraphics[width=0.6\textwidth]{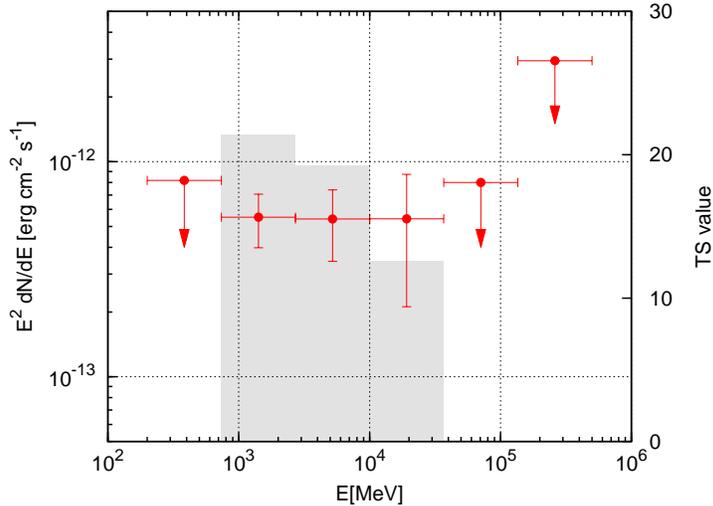}
\caption{The SED of the $\gamma$-ray source, for data between 2014
September 22 and 2017 November 16. The gray shaded regions show the
TS values (right axis).
}
\label{fig:sed}
\end{figure*}

The spectral energy distribution (SED) of the $\gamma$-ray emission for
the data after the supernova explosion is presented in Fig.~\ref{fig:sed}.
The SED gives a flat spectrum which is consistent with that obtained in
the global fit.

\section{Discussion}

\subsection{SDSS J092054.04+504251.5}

A quasi-stellar object, SDSS J092054.04+504251.5, is found to be close to
the $\gamma$-ray source, with an angular separation of $0.045^{\circ}$ from
the best-fit position. The optical spectroscopy of SDSS J092054.04+504251.5
shows strong broad emission lines, suggesting that it is a quasar.
This source is found to lie in the the so-called ``WISE blazar stripe''
in the color - color diagram of the WISE infrared data, and is thus a
candidate blazar which is likely to be a $\gamma$-ray emitter
\citep{2012ApJ...752...61M}. If the $\gamma$-ray source found in this
work is coincident with this quasar, its luminosity is estimated to be
$\sim9\times10^{46}$ erg~s$^{-1}$ adopting a redshift of $z=1.904$.
The $\gamma$-ray luminosity and spectral index are consistent with,
although lie in the edge of, that of {\it Fermi}-LAT flat-spectrum 
radio quasars~\citep{2015ApJ...810...14A}.
However, blazars are typically radio loud. We have searched for
possible radio emission from FIRST and NVSS, and do not find any
counterpart of SDSS J092054.04+504251.5. This may be due to
that this quasar is too distant and the radio surveys are not deep
enough to reveal it. We also search for a possible radio counterpart 
of the $\gamma$-ray source, and do not find any source within its 
95\% error circle. The optical monitorings of SDSS J092054.04+504251.5 
by the Intermediate Palomar Transient Factory gives an average
$R$-band magnitude of 20.44 and a standard deviation of 0.15 between
March 2009 and Januray 2015, which shows no significant variability.
The data after 2015 are unavailable yet, which makes the comparison 
between optical and $\gamma$-ray variabilities impossible. In any case, 
we need more observations of SDSS J092054.04+504251.5 and/or the 
$\gamma$-ray source to establish their possible connection. Currently it 
is unclear whether they are associated with each other, and the association 
of the new $\gamma$-ray variable source with iPTF14hls is possible.

\subsection{Chance coincidence with a background $\gamma$-ray source}

We assume that the surface density distribution of {\it Fermi}-LAT sources 
is uniform, and the probability of observing a source in a particular
position has a Poisson distribution. Then the probability to have
an unrelated source ``associated'' with iPTF14hls is expected to be
\begin{equation}
P_{\rm ch}=1-{\rm exp}\left[-\pi r^2_{\rm eff} \Sigma(>F_{\rm th})\right],
\end{equation}
where $\Sigma(>F_{\rm th})$ is the surface density of {\it Fermi}-LAT 
sources with fluxes higher than $F_{\rm th}$, $r_{\rm eff}$ is an effective
radius which takes into account the position uncertainties of the presumed
{\it Fermi}-LAT counterpart ($\sigma_{\gamma}$) and the target source 
iPTF14hls ($\sigma_{\rm opt}$), as well as the angular distance between 
them, $R_0$ \citep{2002AJ....123.1111B}. For our case, iPTF14hls is well 
located in the optical band, and thus $\sigma_{\rm opt}$ can be neglected.
The effective radius is then $r_{\rm eff}=(R_0^2+4\sigma_{\gamma}^2)^{1/2}$.

\begin{figure*}[!htb]
\centering
\includegraphics[width=0.6\textwidth]{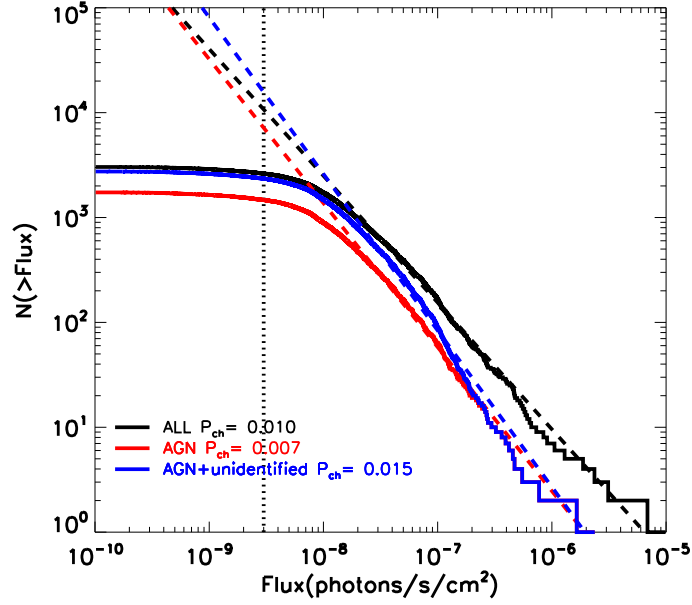}
\caption{The cumulative numbers of sources versus the threshold fluxes 
for {\it Fermi}-LAT 3FGL sources (solid histograms), and the power-law
fitting results of their high-flux trends (dashed lines). 
The black lines are for all the 3FGL sources, red are for AGNs, 
and blue are for AGNs with unidentified sources. 
The vertical dotted line shows the extrapolated $0.1-100$ GeV 
flux of the putative counterpart of iPTF14hls.
}
\label{fig:logNlogP}
\end{figure*}

The cumulative numbers of the 3FGL sources as functions of fluxes are
presented in Fig.~\ref{fig:logNlogP}. In order to avoid the detection
limits, we extrapolate the behaviors for fluxes between $2 \times 10^{-8}$
and $10^{-6}$ photon cm$^{-2}$~s$^{-1}$ to the average flux of the
putative iPTF14hls counterpart converted to the energy range of
$[0.1,100]$ GeV, $3.0 \times 10^{-9}$ photon cm$^{-2}$~s$^{-1}$.
The estimated number of $\gamma$-ray AGNs is 7164, corresponding
to a number density of 0.17 degree$^{-2}$. For $R_0=0.065^{\circ}$ and
$\sigma_{\gamma}=0.045^{\circ}$, the chance coincidence probability is
estimated to be 0.007. We also consider the $N(>{\rm Flux})-{\rm Flux}$
distributions for AGNs plus unidentified sources and all the 3FGL sources,
and get chance coincidence probabilities 0.015 and 0.010, respectively.
Note that the temporal coincidence of the {\it Fermi}-LAT variable source 
with a background source may further decrease the chance coincidence 
probability by a factor of several~(it is roughly the ratio of the whole 
{\it Fermi}-LAT observational time to the time period after the supernova 
explosion).

\subsection{Physical implications of the association with iPTF14hls}

Assuming that the {\it Fermi}-LAT source is associated with the supernova,
we discuss possible physical implications of the $\gamma$-ray emission.
The ejected mass by the supernova iPTF14hls was estimated to be several
tens of solar masses, which corresponds to a total kinetic energy of
\begin{equation}
E_k \sim \frac{M_{\rm ej}v_{\rm sn}^2}{2} \sim 1.8\times10^{52}~{\rm erg}~
\left(\frac{M_{\rm ej}}{50~{\rm M}_{\odot}}\right)~
\left(\frac{v_{\rm sn}}{6000~{\rm km~s^{-1}}}\right)^2,
\end{equation}
where $M_{\rm ej}$ is the ejected mass, and $v_{\rm sn}$ is the velocity
of the ejecta~\citep{2017Natur.551..210A}. The corresponding gas density
of the ejecta is estimated as $n_{\rm gas}\sim\frac{M_{\rm ej}}
{4\pi m_p R^2\Delta R}$, where $m_p$ is the proton mass,
$R \sim v_{\rm sn}t \sim 1.8\times 10^{16}~(t/1~{\rm yr})$ cm is the
radius and $\Delta R$ is the width of the ejecta. Since $\Delta R<R$,
we have $n_{\rm gas}>\frac{3M_{\rm ej}}{4\pi R^3}\approx 2.5\times
10^{9}~(M_{\rm ej}/50~{\rm M}_{\odot})~(t/{\rm yr})^{-3}$ cm$^{-3}$.
The optical emission of iPTF14hls can be fitted with a diluted
blackbody with a temperature of $5000-6000$ K, with a dilution factor
$n_{\rm ph}/n_{\rm bb}$ varying from $\sim 1$ at $t \sim 100$ days and
$\sim 10^{-4}$ at $t \sim 600$ days (inferred from Fig.~4 of
\citealt{2017Natur.551..210A}).

The GeV emission has a total energy of $\sim 10^{51}$ erg. We first
assume that this emission is from the inverse Compton scattering
(ICS)\footnote{The bremsstrahlung emission could be less dominated.
The cooling rate due to ICS is about $10^{-16}$ GeV~cm$^{-3}$
$(E/{\rm GeV})^2~(u/{\rm eV~cm^{-3}})$ with $u$ being the energy
density of photon field, and that due to bremsstrahlung radiation 
in neutral gas is about $10^{-15}$ GeV~cm$^{-3}$ $(E/{\rm GeV})~
(n_{\rm gas}/{\rm cm^{-3}})$. For $t\sim 1$ yr, the energy density
of the photon field is about $3\times10^{10}$ eV~cm$^{-3}$, and the
gas density is about $2.5\times10^9$ cm$^{-3}$. For $E\gtrsim$ a few
GeV, ICS cooling is dominant.}
off the optical photons by energetic electrons accelerated in a certain
site. To boost the optical photons to GeV energies, the Lorentz factors
of electrons are required to be $\gamma_e \sim 10^{4}$. The energy
spectrum of electrons to produce a $E^{-2.0}$ $\gamma$-ray spectrum
is $dN/dE_e \propto E_e^{-3}$ in the slow cooling regime, whose total
power is dominated by low energy particles. Therefore the energy
fraction of high-energy electrons
giving rise to the GeV emission should be very low (e.g., $10^{-4}$
for $E_e>10$ GeV compared with that for $E_e\gtrsim$MeV). In case that
there is a break of the energy spectrum of accelerated electrons below a 
few GeV like it is the case in the Milky Way~\citep{2011A&A...534A..54S},
this fraction could be higher. Furthermore, we need a high enough
acceleration efficiency to convert the kinetic energy to relativistic
electrons. For typical supernovae observed in the Milky Way, the
efficiency of electron acceleration is very low ($10^{-4}\sim
10^{-3}$; \citealt{2014A&A...567A..23Y}). The efficiency for the
case of $\gamma$-ray bursts could be higher. However, we should keep 
in mind that there was no significant decrease of the expansion
velocity of iPTF14hls~\citep{2017Natur.551..210A}, which limits the
conversion fraction from the kinetic energy to the accelerated
particles. Note that here we do not involve jets. If the
$\gamma$-ray emission is collimated in a solid angle, the inferred 
total energy of $\gamma$-rays could be smaller and the required 
acceleration efficiency could be less extreme.

Another scenario to produce $\gamma$-rays is the decay of neutral pions
produced by the hadronic $pp$ collision. To produce the $E^{-2}$
$\gamma$-ray spectrum, the proton spectrum needs to be $dN/dE_p\propto
E_p^{-2}$. The energy conversion efficiency of protons to $\gamma$-rays
depends on the density of target material. The interaction timescale of
protons in the hydrogen gas can be written as
\begin{equation}
t_{\rm pp}\simeq\frac{1}{n_{\rm gas}\sigma_{\rm pp} c}\approx
3\times10^7\left(\frac{n_{\rm gas}}{\rm cm^{-3}}\right)^{-1}{\rm yr}.
\end{equation}
Here $\sigma_{\rm pp} \sim 40$ mb is the total interaction cross
section of $pp$ collision for center-of-mass energy of tens of GeV
\citep{2016ChPhC..40j0001P}. For a gas density higher than
$2.5\times10^9$ cm$^{-3}$ as estimated above, the protons can effectively
convert their energy to $\gamma$-rays, with an efficiency of $\sim33\%$,
the fraction of the neutral pion component. In such a case, the total
energy of protons needs to be $\sim 10^{52}$ erg in order to give the
observed GeV $\gamma$-ray energy (considering also that the energy 
band of the actual $\gamma$-ray emission may be wider than the 
{\it Fermi}-LAT coverage). In this case we may need a
very high conversion efficiency of the supernova kinetic energy to
the accelerated particle energy. Note that the energy conversion
efficiencies obtained from observations of supernova remnants in the
Milky Way are estimated to be about $10\%$
\citep{2010A&A...516A..62A,2017ApJ...843...90X}.
The late time spectroscopic observations by \citet{2017arXiv171200514A}
did find evidence of interactions between the supernova shock and the
circumstellar medium, which supports the scenario of $\gamma$-ray 
production in $pp$ collisions. However, the lack of X-ray and radio 
emission as well as significant deceleration of the ejecta at early time
constrain the possible strong interactions \citep{2017Natur.551..210A}.
Furthermore, even if there were interactions between the supernova
ejecta and the pre-existing dense material, the offset between the 
$\gamma$-ray light curve and the optical one is still a challenge.
We leave detailed modeling of this interaction scenario to future
works.

In both cases discussed above, it could be very difficult for particles
to get accelerated in such a high-density environment without
thermalization. This could be a challenge of the modeling of the
$\gamma$-ray emission from the supernova. In addition, for such a dense
environment, the pair production optical depth of $\gamma$-ray photons
with gas is
\begin{equation}
\tau \sim {\sigma_{\rm pair}M_{\rm ej}\over 4\pi R^2m_p}
\sim 0.15~\left(\frac{M_{\rm ej}}{50~{\rm M}_{\odot}}\right)~
\left(\frac{t}{{\rm yr}}\right)^{-2},
\end{equation}
where $\sigma_{\rm pair}\sim10$ mb is the pair production cross section
for GeV photons \citep{2010SCPMA..53..842W}. The GeV emission, if generated
inside the ejected shell, may be subject to pair production absorbsion by
the material, in particular at early time. This may explain why the
$\gamma$-ray emission appears $\sim300$ days after the explosion.
Here we ignore the Thomson scattering, because the shell is expected to
be optically thin and the radiation field is not sufficient to ionize
the gas~\citep{2017Natur.551..210A}.

\section{Conclusion}

In this work we report the detection of a variable $\gamma$-ray source
which is potentially coincident with the supernova iPTF14hls, using the
{\it Fermi}-LAT data. The $\gamma$-ray source appears $\sim300$ days after
the explosion of iPTF14hls, and is still observable up to $\sim850$ days.
The search for $\gamma$-ray emission from 2008 August 4 to 2015 September
22 result in no detectable emission from this direction. The spectrum of
the source is $\propto E^{-2}$ between 0.2 and 500 GeV, and
the luminosity is $1.0\times 10^{43}~(d/156~{\rm Mpc})^2$ erg~s$^{-1}$.
The isotropic energy of the $\gamma$-ray emission is about $10^{51}$ erg.

There is a quasar, SDSS J092054.04+504251.5, in the error circle of the
{\it Fermi}-LAT source, which is a blazar candidate according to the 
infrared color diagram of WISE. However, no radio counterpart is found
from radio surveys FIRST and NVSS. The lack of multi-wavelength
observations of SDSS J092054.04+504251.5 makes it difficult to conclusively 
address its connection with the $\gamma$-ray variable source.
We also estimate the probability of chance coincidence of the
$\gamma$-ray source with a background source based on the 3FGL catalog,
and results in a low probability of $P_{\rm ch}\lesssim 0.015$.

If the association between the $\gamma$-ray source and iPTF14hls is
real, there are difficulties to model its $\gamma$-ray emission in the
framework of particle acceleration in supernova ejecta produced shocks.
The acceleration efficiency and the energy conversion efficiency of
the accelerated particles to GeV $\gamma$-ray emission need to be high. 
Furthermore, the acceleration of particles in the dense environment of 
the ejecta is also a big challenge. Anisotropic emission from e.g., 
a jet, may be necessary to explain the data \citep{2017arXiv171105180S}.

\acknowledgments
We thank the anonymous referee for helpful suggestions on the paper.
This work is supported by the National Natural Science Foundation of 
China under Grant Nos. 11525313, 11722328, and 11703093, and the 100 
Talents Program of Chinese Academy of Sciences.

\end{document}